\documentclass[runningheads]{llncs}
\usepackage{graphicx}
\usepackage{todonotes}
\usepackage{ulem}
\usepackage{listings}
\usepackage{graphicx}
\usepackage{tcolorbox}
\usepackage{color}
\usepackage{soul}
\usepackage{orcidlink}
\usepackage{caption}
\usepackage{paralist}

\lstdefinelanguage{json}{
    basicstyle=\ttfamily,
    morestring=[b]",
    literate=
     *{0}{{{\color{black}0}}}{1}
      {1}{{{\color{black}1}}}{1}
      {2}{{{\color{black}2}}}{1}
      {3}{{{\color{black}3}}}{1}
      {4}{{{\color{black}4}}}{1}
      {5}{{{\color{black}5}}}{1}
      {6}{{{\color{black}6}}}{1}
      {7}{{{\color{black}7}}}{1}
      {8}{{{\color{black}8}}}{1}
      {9}{{{\color{black}9}}}{1}
      {:}{{{\color{black}{:}}}}{1}
      {,}{{{\color{black}{,}}}}{1}
      {\{}{{{\color{black}{\{}}}}{1}
      {\}}{{{\color{black}{\}}}}}{1}
      {[}{{{\color{black}{[}}}}{1}
      {]}{{{\color{black}{]}}}}{1},
}

\begin{document}

\title{LongEval at CLEF 2025: 
Longitudinal Evaluation of IR Model Performance}
%
%
\author{
Matteo Cancellieri\inst{1} \orcidlink{0000-0002-9558-9772} \and 
Alaa El-Ebshihy\inst{2,3}\orcidlink{0000-0001-6644-2360} \and
Tobias Fink\inst{2,3}\orcidlink{0000-0002-1045-8352} \and \\
Petra Galu\v{s}\v{c}\'{a}kov\'{a}\inst{4} \orcidlink{0000-0001-6328-7131} \and
Gabriela Gonzalez-Saez\inst{5}\orcidlink{0000-0003-0878-5263} \and
Lorraine Goeuriot\inst{5}\orcidlink{0000-0001-7491-1980} \and \\
David Iommi\inst{2}\orcidlink{0000-0002-4270-5709}\and
J\"{u}ri Keller\inst{6} \orcidlink{0000-0002-9392-8646} \and
Petr Knoth\inst{1} \orcidlink{0000-0003-1161-7359} \and
Philippe Mulhem\inst{5} \orcidlink{0000-0002-3245-6462} \and \\
Florina Piroi\inst{2,3} \orcidlink{0000-0001-7584-6439} \and
David Pride\inst{1} \orcidlink{0000-0002-7162-7252} \and
Philipp Schaer\inst{6} \orcidlink{0000-0002-8817-4632}
}
\authorrunning{M. Cancellieri et al.}
%
\institute{
The Open University, Milton Keynes, UK\footnote{Authors ordered alphabetically}
\and
Research Studios Austria, Data Science Studio, Vienna, Austria
\and
TU Wien, Austria
\and
University of Stavanger, Stavanger, Norway
\and
Univ. Grenoble Alpes, CNRS, Grenoble INP\footnote{Institute of Engineering Univ. Grenoble Alpes.}, LIG, Grenoble, France
\and
TH Köln - University of Applied Sciences, Cologne, Germany
}
\maketitle  
\begin{abstract}
This paper presents the third edition of the LongEval Lab, part of the CLEF 2025 conference, which continues to explore the challenges of temporal persistence in Information Retrieval (IR). The lab features two tasks designed to provide researchers with test data that reflect the evolving nature of user queries and document relevance over time. By evaluating how model performance degrades as test data diverge temporally from training data, LongEval seeks to advance the understanding of temporal dynamics in IR systems. The 2025 edition aims to engage the IR and NLP communities in addressing the development of adaptive models that can maintain retrieval quality over time in the domains of web search and scientific retrieval. 

\keywords{Longitudinal Evaluation \and Temporal Persistence \and Temporal Generalisability \and Temporal Change \and Information Retrieval}
\end{abstract}

\setcounter{footnote}{0} 
\section{Introduction}
Information Retrieval (IR) systems are constantly challenged by the evolving search setting~\cite{DBLP:conf/sigir/Dumais14}. The foundational dataset is updated regularly, users develop new information needs, and their perception of relevance varies over time~\cite{DBLP:conf/wsdm/AdarTDE09,ROBERTS2021103865,DBLP:conf/sigir/TikhonovBBOKG13}. These temporal dynamics have strong implications for the aspired goal of maintaining high retrieval effectiveness over time. It is known that search is sensitive to temporal factors~\cite{DBLP:journals/ftir/KanhabuaBN15,liu2024robustneuralinformationretrieval} and that incorporating information from previous points in time can be highly effective~\cite{DBLP:conf/clef/AlexanderFHSHHP24,DBLP:conf/clef/Keller0S24}. Additionally, in modern IR, the systems are updated or retrained often, making them a dynamic component themselves in the evolving search setting. 

While these temporal factors strongly influence retrieval effectiveness, they are often overlooked or abstracted on purpose in conventional evaluations. The results from the previous iterations of the lab showed that the ranking of systems varies over time and that the most effective system is not necessarily also the system that performs the most consistently~\cite{LongEval_CLEF_Conference_2023,LongEval_CLEF_Conference_2024,keller:2024}. This shows how the experimental setup strongly influences the measured effectiveness.

In this third iteration, the LongEval Lab at CLEF (Conference and Labs of the Evaluation Forum\footnote{\url{https://www.clef-initiative.eu/}}) continues to explore the temporal dynamics in IR~\cite{LongEval_CLEF_Conference_2023,LongEval_CLEF_Conference_2024}. This includes the potential and limitations of temporal relevance signals for ranking, the temporal robustness of systems, and novel evaluation methods that factor in time. Thus, this lab sensitises researchers to uncertain and temporally limited validity of conventional evaluation results in IR. Considering the temporal dimension provides a new perspective on search and ultimately leads to a more holistic view on the retrieval problem.
 
This year, the lab provides a unique test bed comprising two evolving test collections. They cover the established retrieval scenarios of Web search and scientific retrieval, which have different goals and distinct dynamics. Participants are invited to submit retrieval runs to two tasks that address these dynamics.
\section{Description of the LongEval 2025 Tasks}
Until 2024, LongEval's information retrieval tasks focused only on retrieving Web documents. In 2025, we enlarge the scope of LongEval as we want to study the potential differences, if any, between two retrieval contexts. The Web retrieval context is a classical Web case, in which very short queries are asked and the very top documents are considered. The scientific search contains potentially longer queries, and users are looking deeper in the result lists. 

Both LongEval tasks use a sequence of datasets collected at different points in time. The (time) distances between two datasets are called ``lags.''  The IR systems that participants design are evaluated on the different lags, computing the differences in evaluation metrics between lags.
\subsection{Task 1: LongEval-WebRetrieval}
This task is a continuation of the Retrieval tasks from the previous two LongEval iterations. It uses evolving Web data to evaluate IR systems longitudinally: the systems are expected to be persistent in their retrieval effectiveness over time. The systems are evaluated on monthly several snapshots of documents and queries (lags), derived from real data acquired from a French Web search engine, Qwant\footnote{Most queries and documents are originally in French. However, English translations are additionally provided as well.}. In this iteration, we evaluate the same IR systems on a sequence of test collections acquired after the last sample of the train collection.

\subsubsection{Lessons learned from the 2024 LongEval edition}
In 2024, the LongEval lab used two test environments, called Lag6 and Lag8. That is, we evaluated IR systems on test data that was 6 and 8 months newer than the data the systems were trained on. 28 teams registered for the second edition of the LongEval Retrieval task, and 14 teams submitted a total of 73 retrieval experiments. The number of teams that submitted is the same as the in the first edition, indicating that the task maintained its popularity.
We observed the following~\cite{LongEval_CLEF_Conference_2024}:
\begin{enumerate}
    \item[\textbf{Approaches:}] Compared to 2023, some participants did use the temporal aspect of the LongEval test collection, incorporating past relevance signals as query reformulation~\cite{DBLP:conf/clef/Keller0S24}. The most effective approaches rely on multi-stage retrieval, using BM25 as a first-stage retrieval and neural-based or LLMs-based models for re-ranking. 
    \item[\textbf{Robustness:}] System rankings were computed with respect to retrieval performances on Lag6 and Lag8 (nDCG scores), and according to the changes in their performance between the two lags (Relative nDCG Drop, RnD, see section \ref{sec:evaluation}). The ranking correlation between the system rankings using nDCGs scores on Lag6 and Lag8 was high, while the ranking correlation between system rankings using RnD scores was low. This points to the fact that systems that are more robust to the evolution of the test collection were not the top-performing ones. This finding is consistent with the findings from Longeval 2023.
    \item[\textbf{Data Preparations:}] As expected in an evolving collection of documents, there are large overlaps in documents and queries between the document snapshots across test collections and between test and train collections. This overlap was not easily identifiable with the released document and query IDs. 
\end{enumerate}

\noindent Based on these observations, in the LongEval 2025 lab we enlarge the training collection with additional snapshots that will allow fine-grained analysis of changes in the data collection from one snapshot to another. Similarly, the test environments will be composed of several consecutive snapshots, allowing for a deeper understanding of data evolution over time. Additionally, the LongEval test collection, in its totality, will be improved in terms of document and query identifiers, and its description.
\subsubsection{Data.}
%

The 2025 dataset includes all data from the 2023 and 2024 editions, along with newly added, previously unreleased months. The training dataset consists of 18 million French documents (June 2022 - February 2023, translated to English) and 9,000 queries with computed relevance assessments based on a simplified Dynamic Bayesian Network (sDBN) Click Model~\cite{Chapelle2009-rg,Chuklin2015}, acquired from real users of the French Qwant search engine.
The test collection spans 7 months of data (March 2023 - August 2023). Each month can be seen as a snapshot (lag). Each of these test collections is similar in structure to the train set, except that they do not contain any relevance assessments. Participants are expected to submit runs for each lag, using the same system trained only on the training dataset. Additionally, human-annotated data from the previous iteration will be used to evaluate system performance.
The total data for this task will be composed of 30 million documents and 15,000 queries, provided by Qwant\footnote{Qwant search engine: \url{https://www.qwant.com/}}. Each document set will have a release time stamp, with the first set (in chronological order) being the training data.

\subsection{Task 2: LongEval-SciRetrieval}

The second task of the LongEval 2025 Lab is similar to the first task, and aims to examine how IR systems' effectiveness changes over time, when the underlying document collection changes, where the documents are scientific publications.  The documents that will make the dataset for this task are acquired from the CORE\footnote{CORE (COnnecting REpositories) https://core.ac.uk/} collection of scholarly documents. 
To our knowledge, CORE \cite{knoth2023core} is currently the largest aggregated collection of Open Access full text scholarly documents. CORE provides a range of services built on top of this content and these services are currently used by over 30 million unique users each month. CORE Search provides a web UI for users to query the entire database of scholarly documents. This service registers over one million searches each month.

\begin{figure}[h!]
    \centering
    \includegraphics[width=0.7\textwidth]{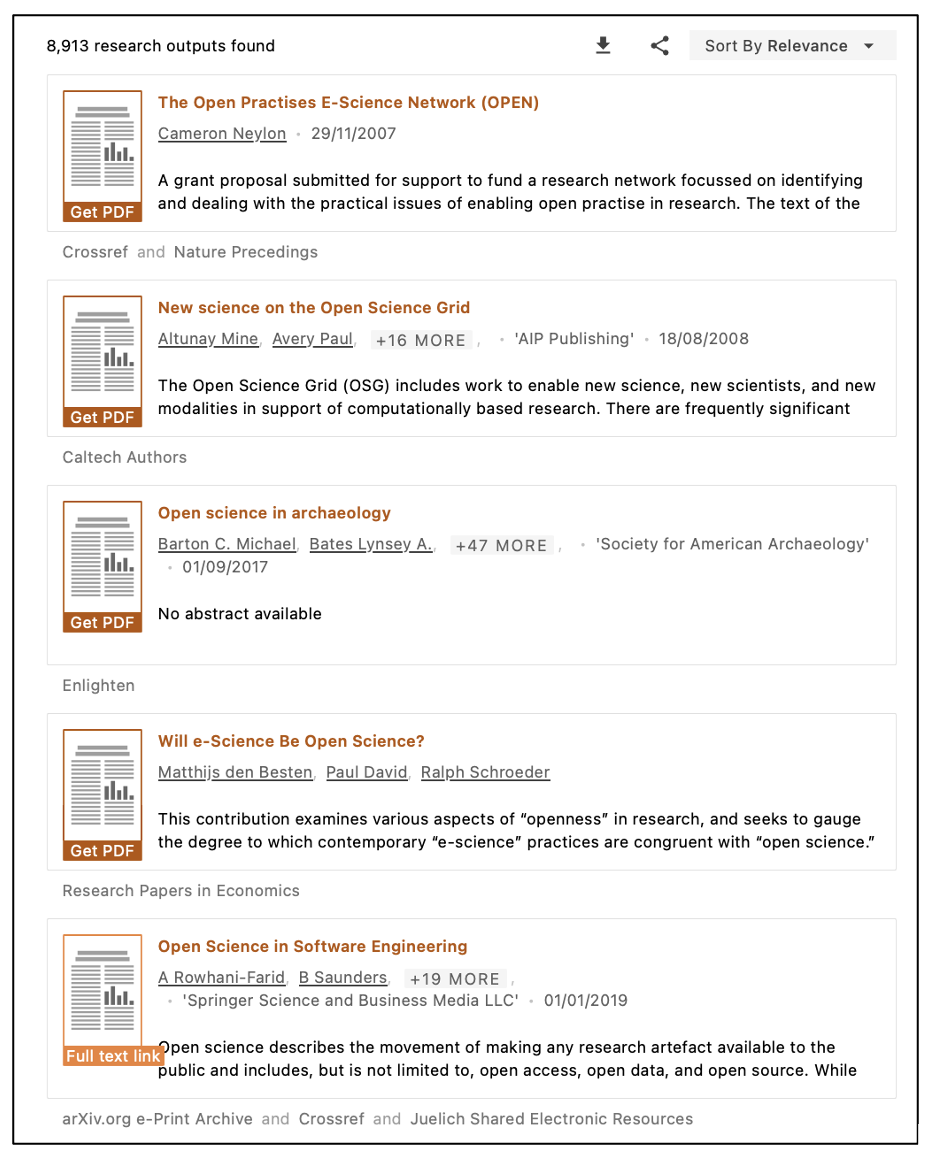}
    \caption{Sample result from CORE for a search for ``open science''}
    \label{fig:open_science}
\end{figure}

As can be seen from the sample results shown in Figure~\ref{fig:open_science} for the query ``open science'', the user has multiple options in terms of where to click for each individual search result: the PDF link (left hand image), the paper title (in brown color) and the author(s) name (underlined). 
Similarly to Task 1, we will use the click information to create relevance assessments for the test collection.

For compiling the dataset for the LongEval-SciRetrieval task, we create a specific pipeline for capturing user queries, the results of these queries, and the user interactions with the displayed results from the CORE Apache server logs. We then process the data to remove any traffic that was generated by bots so only searches conducted by human users remain. Using this pipeline, the dataset for this task is extracted and consists of two main components that contain both the search and click information:
\begin{enumerate}
    \item[\textbf{Search Information}] (see example~\ref{ex:search}) includes i) unique (anonymous) identifiers for individual user session; ii) search query; iii) returned results\footnote{For LongEval we only consider the first ten results.} ;
    \item[\textbf{Click Information}] (see example~\ref{ex:click}) records, for each click, i) a unique (anonymous) identifier for individual user session; ii) the link that was clicked in the results list; iii) the position of clicked link in results list.
\end{enumerate}
Since this is the first time this task is organized, the number of dataset lags is lower than those used in the first task. We aim to release two training datasets and one or two test datasets.



\newpage

\begin{lstlisting}[label={ex:search},language=json, caption={Search query and returned results}]
{ 
  "search_id": "e5385afdaedcc11f9a6ba092cb613f27", 
  "query": "inclusive codesign",
  "serp": [
    "https://core.ac.uk/works/156973302",
    "https://core.ac.uk/works/8080300",
    "https://core.ac.uk/works/45177790",
    "https://core.ac.uk/works/148924361",
    "https://core.ac.uk/works/149405922",
    "https://core.ac.uk/works/149554048",
    "https://core.ac.uk/works/15034633",
    "https://core.ac.uk/works/150382029",
  ],
  "date": "2024-12-03T07:11:11.000Z"
}
\end{lstlisting}

\begin{lstlisting}[label={ex:click},language=json, caption={Unique click information}]
{ 
  "search_id": "e5385afdaedcc11f9a6ba092cb613f27", 
  "url": "https://core.ac.uk/works/156973302",
  "serp": "0", 
  "date": "2024-12-03T07:11:11.000Z" 
}
\end{lstlisting}

\section{Evaluation}
\label{sec:evaluation}
Since the two retrieval tasks are very similar in design, differing in the type of data provided to the users (Web documents vs. scientific publications), the evaluation is, conceptually the same. Namely, the submitted runs will be mainly evaluated in two ways:
\begin{enumerate}
    \item \textbf{nDCG} scores calculated on each lag test set provided for the sub-tasks. Such a classical evaluation measure is consistent with Web search, for which the discount emphasises the ordering of the top results.
    \item \textbf{Relative nDCG Drop (RnD)} measured by computing the difference between nDCG values between different lag datasets. Such values will allow to check the robustness of systems against the evolution of the data.
\end{enumerate}
These measures assess the quality of systems and also their robustness against the data (queries/documents) evolution along time: a system that has good results using nDCG, and also good results according to the RnD measure is considered to be able to cope with the evolution over time of the Information Retrieval collection.

\section{LongEval Timeline}
%
Information and updates about the LongEval Lab, and the submission guidelines, will be communicated mainly through the lab's website\footnote{\url{https://clef-longeval.github.io}}.
The training data release for both tasks is scheduled for February 2025, and the test data for end of March 2025. In concordance with the CLEF schedule, the participant submission deadline is planned for May 2025, with the evaluation results to be released in June 2025.
As in the previous iterations, we invite participants to the LongEval workshop to be organized as part of the CLEF 2025 conference. The workshop is open to researchers interested in the temporal persistence of IR models, and we welcome submissions that are not part of the shared task but deal with this topic.
\section*{Acknowledgements}
This work is supported by the ANR Kodicare bi-lateral project, grant ANR-19-CE23-0029 of the French Agence Nationale de la Recherche, and by the Austrian Science Fund (FWF, grant I4471-N). This work has been using services provided by the LINDAT/CLARIAH-CZ Research Infrastructure (https://lindat.cz), supported by the Ministry of Education, Youth and Sports of the Czech Republic (Project No. LM2023062).

\bibliographystyle{splncs04}
\bibliography{longeval}

\end{document}